\documentclass[11pt]{article}
\usepackage{blois,epsfig}

\def\be{\begin{equation}}
\def\ee{\end{equation}}
\def\bea{\begin{eqnarray}}
\def\eea{\end{eqnarray}}

\begin{document}
\vspace*{4cm}
\title{GALACTIC HYDRAULIC DROP?}

\author{STEEN H. HANSEN}

\address{University of Zurich, Winterthurerstrasse 190,
CH-8057 Zurich,
Switzerland}

\maketitle\abstracts{Galaxies and galaxy clusters are observed to have
a rather non-trivial radial behaviour.  The observations show that the
radial profiles change from one power-law profile near the centre to
another power-law profile in the outer region.  We present a simple
explanation for this complex behaviour by finding the analytical
solutions to the governing hydrodynamic equations. We see that the
origin of this complexity is the collisional nature of the baryonic
plasma, possibly related to a turbulence-enhanced viscosity.}

\section{Introduction}

Large gaseous baryonic structures such as galaxies and galaxy clusters
have been known and observed for many years. A characteristic
behaviour is that the radial profiles, e.g.\ of surface brightness or
electron density, often have the complex behaviour that they follow
one power-slope, $\alpha$, in the inner part, and another power-slope,
$\beta$, in the outer part beyond a characteristic radius, $r_0$,
\begin{equation}
A = \frac{A_0 }{\left( \frac{r}{r_0}\right)^\alpha
\left(1 + \frac{r}{r_0}\right)^\beta} \, ,
\label{eq:profile}
\end{equation}
and this transition is frequently observed to be rather sharp.  This
behaviour is often fit by observers by simple phenomenological
profiles like eq.~(\ref{eq:profile}) which is composed of just such
two power-laws.  These observations include spiral galaxies (e.g.
using WFPC2 data~\cite{carollo,gebhardt}) and clusters of galaxies
(e.g. using Chandra data~\cite{lewis}).  However, there is little (if
any) theoretical guidance for the use of such profiles.  Here we
attempt a derivation of this complex behaviour.

Traditionally one would expect that two different power-slopes must be
related to different physics, in particular if the transition is
sharp. Surprisingly enough, for the radial density profile this does
not have to be the case. We will show that the governing equations
have exactly two solutions, which imply that the inner and outer
density profiles may choose different solutions, and hence quite
generally will be different (see ref.~[4] for details).

\begin{figure}
\center{
\psfig{figure=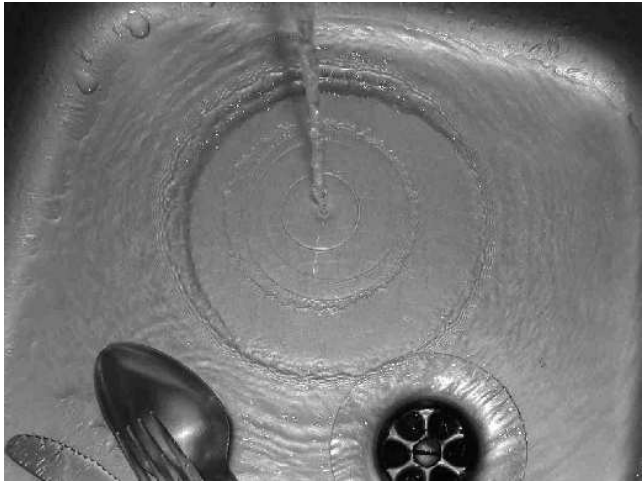,height=2.5in}}
\caption{The {\em circular hydraulic jump}, as observed in any kitchen sink.
The water chooses one solution inside the jump and another solution
outside the jump. We show that galaxy clusters behave in a similar
manner, and that the density profile (or surface brightness)
therefore is expected to exhibit a break at some characteristic
radius, providing us with a {\em galactic hydraulic drop}.\label{fig:hydrjump}}
\end{figure}

In the field of hydrodynamics cases are known where a given set of
equations have two solutions, and that Nature chooses to use {\em
both} solutions simultaneously. One well-known example is the {\it
hydraulic jump}, which is a centimetre large ring, observed in any
kitchen sink when the water flows out radially after hitting the sink
(see Figure~1).  The water in the inner few centimetre follows one
solution, and outside the jump the water follows another
solution\cite{hansen97,bohr}. It turns out that in a similar manner
the density profile in the inner part of e.g.\ a relaxed galaxy
cluster chooses one solution, whereas the outer part of the same
cluster chooses another solution.

\section{Solving the Navier-Stokes equations analytically}

The behaviour of any collisional gas or fluid is fully determined by
the Navier-Stokes (N-S) equations.
Baryons often have sufficient collisions to be described by the N-S equations, 
e.g. in a typical intra-cluster gas the 
equilibration timescale is
about $10^7$ years, with mean free path of tens of kpc compared to
radii of few Mpc.

In this proceeding we will for simplicity consider a stable
spherical cluster of galaxies, where the system has picked out an
orientation in space, such that all the gas is moving only in the
$\Theta$-direction. Thus we have $v_r = v_\phi=0$.  Here we use
notation where $r$ is the radial coordinate, $\Theta$ is the angle in
the $xy$-plane, and $\phi$ is the angle from the $z$-axis. One
must keep in mind that by considering the N-S equations we are taking
a fluid approach which implies that we are following a fluid element,
and this basically corresponds to averaging over all the particles
moving through that fluid element.  For the $\Theta$-velocity we
consider the simple form
\begin{eqnarray}
v_\Theta &=& v_\alpha \, \left( \frac{r}{r_\alpha}  \right)^\alpha \,
{\rm sin}\phi \, , \label{eq:v}
\end{eqnarray}
where $\alpha$ is the constant to be determined first, $v_\alpha$
and $r_\alpha$ are unknown constants, with the physical interpretation
that $r_\alpha$ is a characteristic transition radius,
and $v_\alpha$ is the velocity of the fluid element at that radius.

{\bf The first N-S equation.} 
The first N-S equation becomes very simple with the assumed form of
the velocities
\begin{equation}
0 = \nu \, \left[ \nabla^2 v_\Theta  
- \frac{v_\Theta}{r^2 {\rm sin}^2 \phi} \right] \, ,
\label{eq:vis}
\end{equation}
where $\nabla^2$ is the scalar Laplacian, and $\nu$ is the kinematic
viscosity. For now all that matters is the existence of a non-zero
viscosity, so the absolute magnitude (and even radial dependence) is
not important for the results. Certainly baryons have 
  non-zero viscosity, however the viscosity could also arise from
  turbulence, in which case it could be very large~\cite{ll87}, 
$\nu_{\mbox{turb}}
  \sim l \Delta v$, where the dimension $l$ is the size of
  the turbulent eddies, and $\Delta v$ is the velocity dispersion.
Numerical simulations~\cite{normanbryan} 
have shown that such turbulence indeed exists
in galaxy clusters with $\Delta v \sim 300 - 600$ km/sec, and 
$l \sim 100 -500$ kpc, leading to a very large (turbulence enhanced)
viscosity.

When we use the form for $v_\Theta$ in eq.~(\ref{eq:v}), then
eq.~(\ref{eq:vis}) has exactly 2 solutions
\begin{equation}
\alpha = 1 \, \, \, \, \, {\rm or} \, \, \, \, \, 
-2 \, . \label{eq:plus}\\
\end{equation}
Thus, looking at eq.~(\ref{eq:v}), it is clear that the solution with
$\alpha = -2$ is divergent for $r \rightarrow 0$, and we will
therefore refer to this solution as the {\em 'outer solution'}, and
similarly, the solution with $\alpha=+1$ is inconsistent for large
radii, and we will refer to this as the {\em 'inner solution'}. Hence
the general flow pattern changes from the inner to the outer region,
and one may therefore expect to find different density profiles in the
central and outer regions.

{\bf The second N-S equation.} 
We can now use the next N-S equation to extract the asymptotic radial
density profiles. Also this equation is very simple
\begin{equation}
-\frac{v_\Theta^2}{r} = -\frac{1}{\rho}\frac{\partial P}{\partial r}
- \frac{M(r) G}{r^2} \label{eq:vr} \, ,
\end{equation}
where $\rho$ is the radially dependent density, $P$ is the pressure,
$G$ is the gravitational constant, and $M(r)$ is the mass within the
radius $r$.  We assume that the pressure and density are related
through $P = P_\alpha\, (\rho/\rho_\alpha)^\gamma$, where $P_\alpha$
and $\rho_\alpha$ are the unknown pressure and density at $r_\alpha$.
We take a monatomic gas with $\gamma=5/3$.  Let us consider densities
of the form
\begin{equation}
\rho(r) = \rho_\alpha \, \left( \frac{r}{r_\alpha}\right)^\beta \, ,
\label{eq:rho}
\end{equation}
such that the parameter $\beta$ determines the density profile.  It is
worth emphasizing that it is exactly this $\beta$ which we are trying
to find. 

The last (gravitational) term including $M(r)$ depends on the given
system we are considering. If the mass is dominated by a point
gravitational source (e.g. a central black hole (BH)), then it goes
like $M(r) G/r^2 \sim r^{-2}$.  If the mass is dominated by the matter
density, then it goes like $M(r) \sim \int \rho(r) dV$, with $dV$ the
volume element.  For spherical solutions this gravitational term thus
goes like $r^{\beta+1}$ with $\beta$ from eq.~(\ref{eq:rho}).

It is now straight forward to solve eq.~(\ref{eq:vr}) in the inner
and outer regions. 
The $\alpha=1$ is the inner solution, for which  we find $\beta = -6$
when the baryons dominate the mass. With BH dominance one has
$\beta = -3/2$, and 
if dominated by 
another spherical distribution (which probably should arise 
from dark matter (DM)) 
with profile $\tilde \beta_s$, then we find
$\beta = 3/2 \, (\tilde \beta_s + 2)$.
For a different polytropic index, $1< \gamma <5/3$,
the coefficient changes from $-3/2$ to $(\gamma -1)^{-1}$.
Thus, if the DM has a slope of $\tilde \beta_s = -2$ (as expected from
adiabatic contraction), then the baryons
develop a core, $\beta = 0$. 

In the outer region ($\alpha = -2$) we find $\beta = -6$. If a BH
dominates then $\beta = -3/2$, and again if a another spherical 
distribution (DM)  dominates then $3/2 \, (\tilde \beta_s + 2)$.
E.g.\ a DM slope of $-3$ leads to $\beta = -3/2$.
One should, however, keep in mind that in the outer
region there are possibly not sufficient collisions to assure the
validity of the N-S equations, so those simple solutions should not
be trusted too much.

\section{How to test these findings}
Our main finding is that a transition from the inner to the outer
region generally exists. This is observationally well
established~\cite{carollo,gebhardt,lewis}.  We have thereby provided
theoretical support for the use of phenomenological profiles like
eq.~(\ref{eq:profile}).  One can now take a further step and test the
actual numbers we find in our simplified treatment.

The findings for BH domination always give $\beta=-3/2$, which is just
what numerical simulations find~\cite{young80}.  The actual density
profiles can be observed in different ways.  X-ray observations of the
luminosity in various bands give the electron density of the plasma as
a function of radius.  E.g. for the relaxed cluster A2029 the outer
baryonic profile of $\beta =-1.62$ was found~\cite{lewis}. In the
future the Sunyaev-Zeldovich effect will directly provide a measure of
the radial electron density~\cite{hansen,aghanim}, which will probe
large cluster radii since the SZ effect is proportional to $n_e$
whereas X-ray observations are proportional to $n_e^2$.  Surface
brightness from radio observations of $HI$ and molecular gas can in
principle determine the baryon profile~\cite{corbelli}. 
For details and discussion see ref.~[4].
We are looking
forward to doing a more detailed analysis.

\section{Conclusions}
We have presented an explanation for the origin of the complex radial
structure of galaxy clusters. Specifically, we have shown that the
density profiles generally are expected to make a transition from one
power-slope in the inner to another power-slope in the outer region
\begin{equation}
\rho_{\rm gas} (r) = \frac{\rho(0)}{r^{\beta_1} 
( 1 + r )^{\beta_2}}  \, .
\end{equation}
The physical origin of this complexity is the collisional nature of
the baryonic plasma, and we speculate that it may be related to a
turbulence-enhanced viscosity.

\section*{Acknowledgments}
It is a pleasure to thank Joachim Stadel for collaboration, and 
the Tomalla foundation for financial support.

\end{document}